\documentclass[prc,preprint]{revtex4}

\usepackage{graphicx}

\begin{document}
\title{\bf Computation of the structure of magnetized strange quark star }

\author{\bf G.H. Bordbar $^{1,3}$\footnote{Corresponding author}
\footnote{E-Mail: bordbar@physics.susc.ac.ir} and A. R. Peivand
$^2$}
\affiliation{ $^1$Department of Physics, Shiraz University,
Shiraz 71454, Iran\footnote{Permanent address},\\
$^2$Department of Physics, Tafresh University, Tafresh, Iran\\
$^3$Research Institute for Astronomy and Astrophysics of
Maragha,\\
P.O. Box 55134-441, Maragha, Iran}

\begin{abstract}
In this work, we have calculated some properties of the spin
polarized strange quark matter (SQM) in a strong magnetic field at
zero temperature using the MIT bag model.
We have shown that the equation of state of spin polarized SQM is
stiffer than that of the unpolarized case.
We have also computed the structure properties of the spin
polarized strange quark star (SQS) and have found that the
presence of magnetic field leads to a more stable SQS compared to
the unpolarized SQS.
\end{abstract}
\maketitle

\section{Introduction}
Strange quark stars (SQS) are those which are built mainly from
self bound strange quark matter (SQM).
The surface density of SQS is equal to the density of SQM at zero
pressure ($\sim 10^{15}\ g/cm^3$), which is fourteen orders of
magnitude greater than the surface density of a normal neutron
star. The central density of these stars is about five times
greater than their surface density
\cite{haensel,glendening,weber,10}.
The existence of SQS which are made of SQM was first proposed by
Itoh \cite{a} even before the full developments of QCD.
Later Bodmer \cite{b} discussed the fate of an astronomical object
collapsing to such a state of matter.
In 1970s, after the formulation of QCD, the perturbative
calculations of the equation of state of the SQM was developed,
but the region of validity of these calculations was restricted
to very high densities \cite{collins}.
The existence of SQS was also discussed by Witten \cite{c} who
conjectured that a first order QCD phase transition in the early
universe could concentrate most of the quark excess in dense quark
nuggets. He suggested that the true state of matter was SQM.
Witten proposal was that the SQM composed of light quarks is more
stable than nuclei, therefore  SQM can be considered as the
ground state of matter.
SQS would be the bulk SQM phase consisting of almost equal numbers
of up, down, and strange quarks plus a small number of electrons
to ensure the charge neutrality.
A typical electron fraction is less than $10^{-3}$ and it
decreases from the surface to the center of SQS
\cite{haensel,glendening,weber,10}.
SQM would have a lower charge to baryon ratio compared to the
nuclear matter and can show itself in the form of SQS
\cite{c,d,e,f}.

The collapse of a massive star may lead to the formation of a SQS.
A SQS may be also formed from a neutron star and is denser than
the neutron star \cite{2}.
If sufficient additional matter is added to a SQS, it will
collapse into a black hole. Neutron stars with masses of $1.5-1.8
M_\odot$ with rapid spins are theoretically the best candidates
for conversion to the SQS. An extrapolation based on this
indicates that up to two quark-novae occur in the observable
universe each day.
Besides, recent Chandra observations indicate that objects RX
J185635-3754 and 3C58 may be bare SQS \cite{prakash}.


It is known that the compact objects such as the neutron stars,
pulsars, magnetars, and strange quark stars are under the
influence of the strong magnetic field, which typically is about
$10^{15}-10^{19}\ G$ \cite{kouv1, kouv2, haensel, glendening,
weber,10}.
Therefore, in astrophysics, it is of special interest  to study
the effect of strong magnetic field on SQM properties which can be
found in the core of neutron stars and also in the SQS. We note
that in the presence of magnetic field, the conversion of neutron
stars to bare quark stars can not take place unless the value of
magnetic field exceeds $10^{20}\ G$ \cite{chak}.

Recently, we have calculated the structure of unpolarized SQS at
zero temperature \cite{nurafshan} and finite temperature
\cite{zamani}. In this article, we focus on SQS which is purely
composed of the spin polarized SQM, and investigate the effects of
strong magnetic field on different properties of such an star. In
section 2, we study the spin polarized SQM in the absence and
presence of the strong magnetic field. In section 3, by
numerically solving the Tolman-Oppenhaimer-Volkoff equation, we
obtain the structure properties of the spin polarized SQS.
Moreover, we discuss the stability of spin polarized SQS.

\section{Energy calculation for the spin polarized SQM }
We consider the spin polarized SQM composed of $u$, $d$, and $s$
quarks with spin up ($+$) and down ($-$). We denote the number
density of quark $i$ with spin up by $\rho^{(+)}_{i}$, and spin
down by $\rho^{(-)}_{i}$. We introduce the polarization parameter
$\xi_i$ by
\begin{equation}\label{equ1}
\xi_{i}=\frac{\rho^{(+)}_{i}-\rho^{(-)}_{i}}{\rho_{i}},
\end{equation}
where $0\leq\xi_{i}\leq1\,$ and
$\rho_i=\rho^{(+)}_{i}+\rho^{(-)}_{i}$. Under the conditions of
beta-equilibrium and charge neutrality, we get the following
relation for the number density,
\begin{equation}\label{equa11}
\rho=\rho_{u}=\rho_{d}=\rho_{s},
\end{equation}
where $\rho$ is the total baryonic density of the system.

Now, we calculate the energy density of spin polarized SQM.
To calculate the total energy of spin polarized SQM, we use MIT
bag model in which the total energy is the sum of kinetic energy
of quarks plus a bag constant ($B_{bag}$) \cite{chodos}. The bag
constant $B_{bag}$ can be interpreted as the difference between
the energy densities of the noninteracting quarks and the
interacting ones. Dynamically it acts as a pressure that keeps the
quark gas in constant density and potential. In MIT bag models,
different values are considered for the bag constant such as $55$
and $90\ \frac{MeV}{fm^3}$ .
We calculate the energy density of SQM in the absence and presence
of the magnetic field in the following two separate sections.

\subsection{Energy density of spin polarized SQM in the absence of magnetic field}
The total energy of the spin polarized SQM in the absence of
magnetic field ($B=0$) is given by

\begin{equation}\label{eq6}
\varepsilon_{tot}^{(B=0)} = \varepsilon_u + \varepsilon_d +
\varepsilon_s + {B_{bag}},
\end{equation}
where $\varepsilon_i$ is the kinetic energy per volume of quark
$i$,
\begin{equation}
\varepsilon_i=\sum_{p=\pm}\ \sum_{k^{(p)}}
\sqrt{m_i^2c^4+\hbar^2{k^{(p)}}^2c^2} .
\end{equation}
We ignore the masses of quarks $u$ and $d$, while we consider
$m_{s}=150\, MeV$ for quark $s$. After doing some algebra,
supposing that $\xi_{s}=\xi_{u}=\xi_{d}=\xi$,  we get the
following relation for the total energy of the spin polarized SQM,

\begin{eqnarray}\label{etotabs}
\varepsilon^{(B=0)}_{tot}&=&\frac{3}{16\pi^{2}\hbar^{3}}{\large\sum_{p=\pm}}
\left[\frac{\hbar}{c^{2}}\,k_{F}^{(p)}E_{F}^{(p)}\left( 2
\hbar^{2}k_{F}^{(p) 2}c^{2}+
m^{2}_{s}c^{4}\right)-m^{4}_{s}c^{5}\ln (\frac{\hbar
k_{F}^{(p)}+ E_{F}^{(p)}/c}{m_{s} c})\right]\nonumber\\
&+&\frac{3\,\hbar c
\pi^{2/3}}{4}\,\rho^{4/3}\left[(1+\xi)^{4/3}+(1-\xi)^{4/3}\right]+B_{bag},
\end{eqnarray}
where
\begin{equation}
k_{F}^{\pm}=(\pi^{2}\rho)^{1/3}(1\pm\xi)^{1/3},
\end{equation}
and
\begin{equation}
E_{F}^{\pm}=\left(\hbar^{2}k_{F}^{(\pm) 2}c^{2}+
m_s^{2}c^{4}\right)^{1/2}.
\end{equation}

\begin{figure}
\includegraphics[width=12cm]{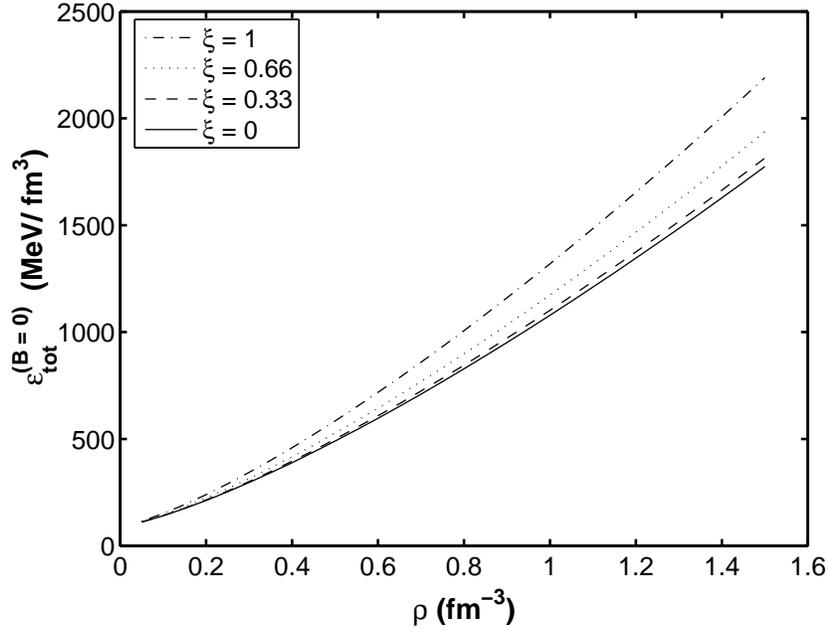}
\caption{The total energy density of spin polarized SQM as a
function of the density ($\rho$) at different values of the
polarization parameter ($\xi$) in the absence of magnetic field.}
\label{1}
\end{figure}

In Fig. \ref{1}, we have plotted the total energy density of spin
polarized SQM as a function of the density for different values of
the polarization ($\xi$) in the absence of magnetic field.
Fig. \ref{1} shows that the energy is an increasing function of
the density, however the increasing rate of energy versus density
increases by increasing  polarization.
For each density, we see that the energy of spin polarized SQM
increases by increasing polarization, specially at high densities.

\begin{figure}
\includegraphics[width=14cm]{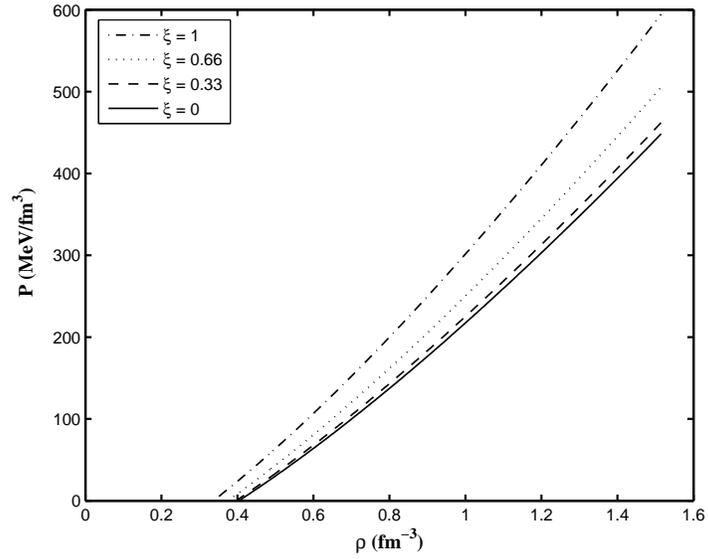}
\caption{As Fig. \ref{1} but for the equation of state of spin
polarized SQM.}\label{2}
\end{figure}

For the spin polarized SQM, we can also calculate the equation of
state (EoS) using the following relation,
\begin{equation}\label{eos}
P(\rho)=\rho\frac{\partial\varepsilon_{tot}}{\partial\rho}-\varepsilon_{tot},
\end{equation}
where $P$ is the pressure and $\varepsilon_{tot}$ is the energy
density which in the absence of magnetic field, is obtained from
Eq. (\ref{etotabs}).
In Fig. \ref{2}, we have shown the pressure of spin polarized SQM
as a function of the density for various values of the
polarization parameter in the absence of magnetic field. We see
that for a given density, the pressure increases by increasing
polarization. This shows that the EoS of spin polarized SQM is
stiffer than that of the unpolarized case.
From Fig. \ref{2}, it can be seen that by increasing polarization,
the density corresponding to zero pressure  takes lower values.
\subsection{Energy density of spin polarized SQM in the presence of magnetic field }
In this section, we consider the spin polarized SQM which is under
influence of a strong magnetic field (${\bf B}$). For this system,
the contribution of magnetic energy is  $E_{M}=-{\bf M\cdot B}$.
If we consider the magnetic field along  $z$ direction,  the
contribution of magnetic energy of the spin polarized SQM is given
by
\begin{equation}
E_{M}=-\sum_{i=u,d,s} M^{(i)}_{z}B,
\end{equation}
where $M^{(i)}_{z}$ is the magnetization of system corresponding
to particle $i$ which is given by
\begin{equation}
M^{(i)}_{z}=N_{i}\mu_{i}\xi_i.
\end{equation}
In the above equation, $N_{i}$ and $\mu_{i}$ are the number and
magnetic moment of particle $i$, respectively. By some
simplification, the contribution of magnetic energy density of the
spin polarized SQM, $\varepsilon_{M}=\frac{E_{M}}{V}$, can be
obtained as follows,
\begin{equation}\label{equ2}
\varepsilon_{M}= -\sum_{i=u,d,s}\rho_{i}\mu_{i}\xi_i B.
\end{equation}
Consequently, the total energy density of spin polarized SQM in
the presence of magnetic field can be written as
\begin{eqnarray}\label{equa121}
\varepsilon^{(B)}_{tot}&=&\varepsilon_{tot}^{(B=0)} +
\varepsilon_{M}.
\end{eqnarray}

\begin{figure}
\includegraphics[width=14cm]{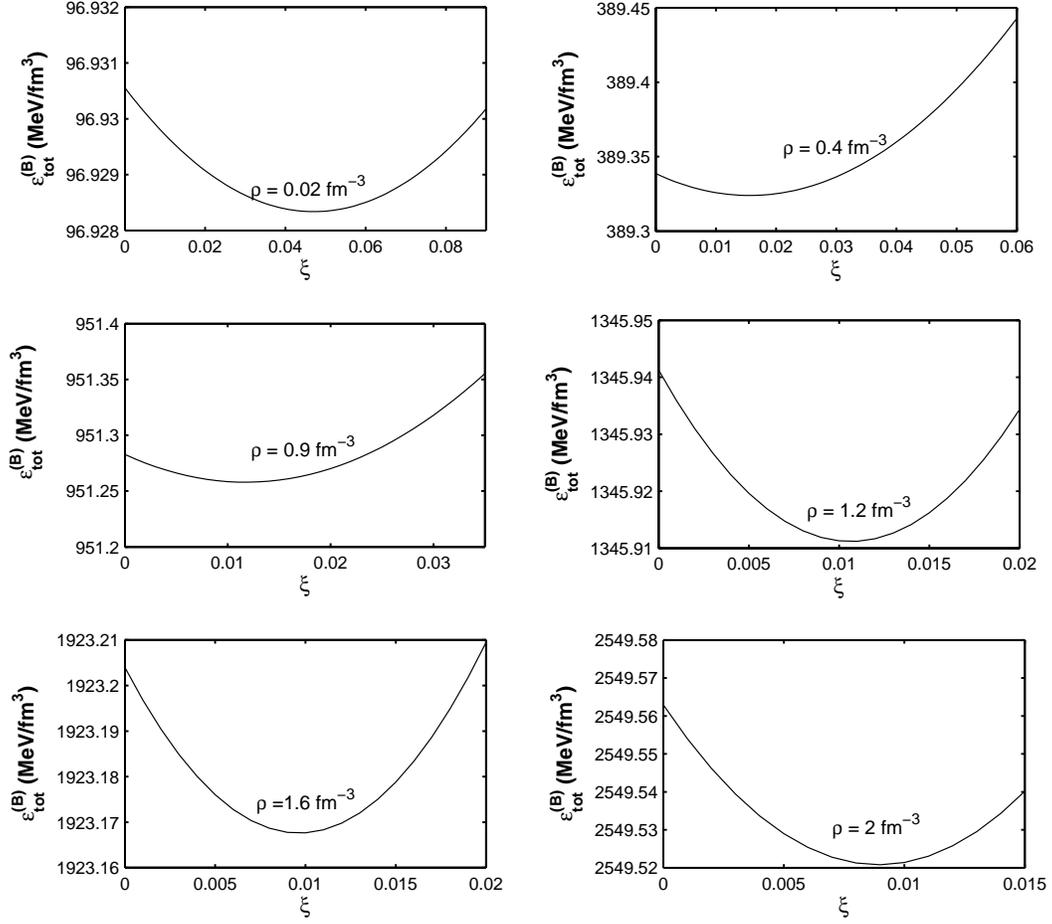}
\caption{ The total energy density of polarized SQM as a function
of the polarization parameter ($\xi$) for $B=5\times 10^{18}\ G$
at different densities ($\rho$). }\label{3}
\end{figure}

\begin{figure}
\includegraphics[width=12cm]{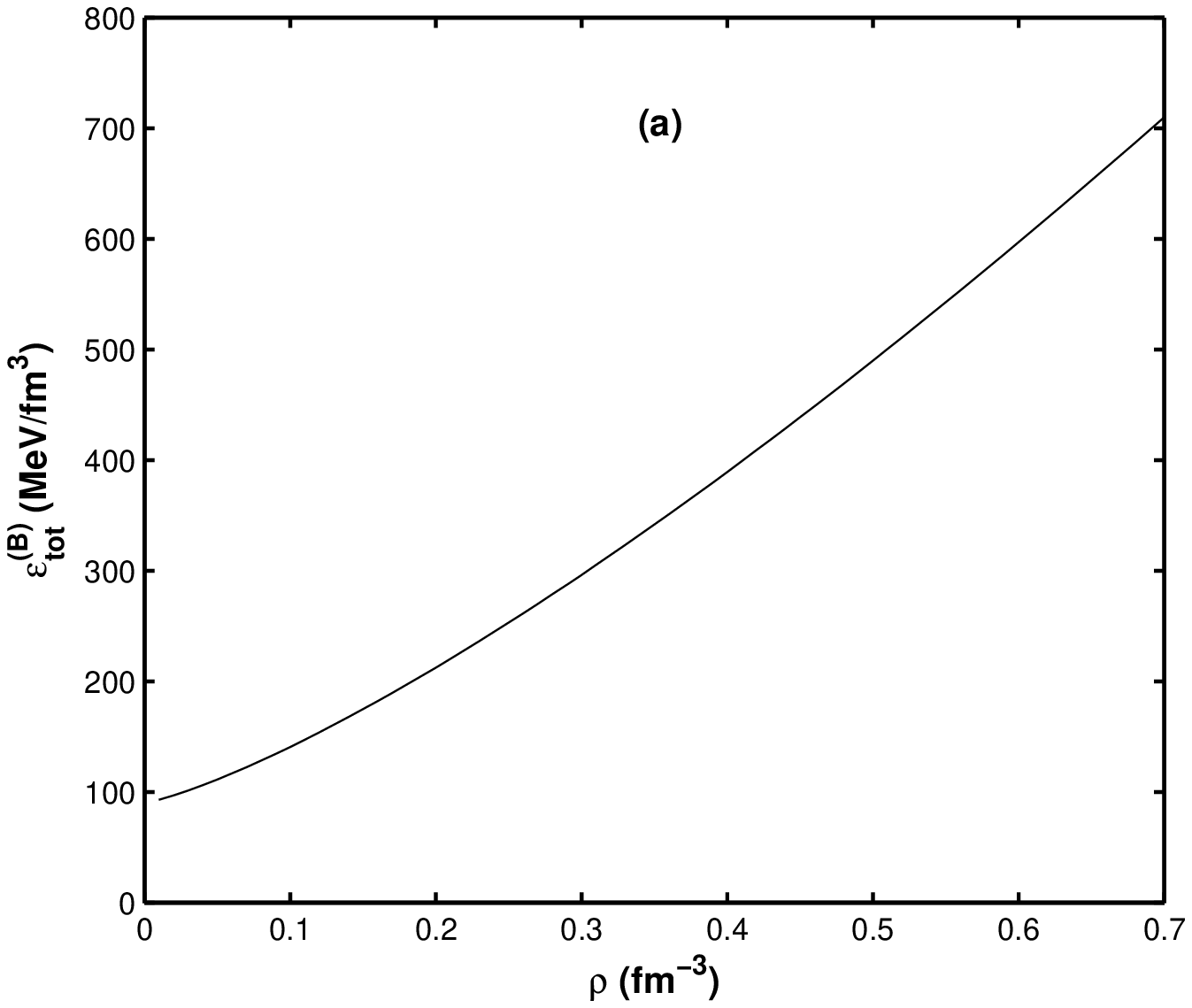}
\includegraphics[width=12cm]{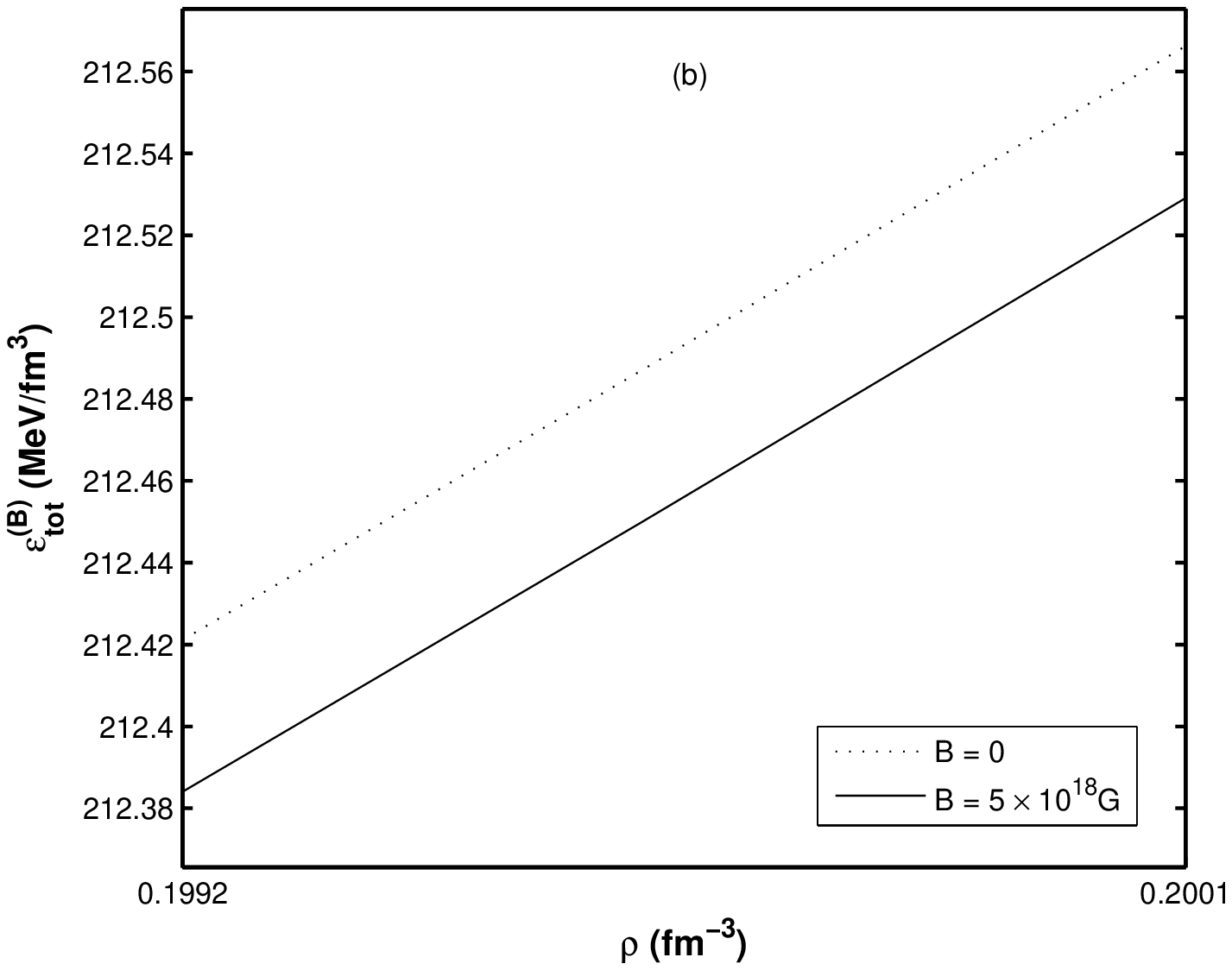}
\caption{(a) The total energy density of spin polarized SQM versus
the density ($\rho$) at $B=5\times 10^{18} G$. (b) Comparison
between the total energy for two cases of $B=5\times 10^{18}\ G$
and $B=0$.} \label{4}
\end{figure}

In Fig. \ref{3}, we have shown the total energy density of the
spin polarized SQM as a function of the polarization parameter
($\xi$), for $B=5\times10^{18} G$ at various densities. From Fig.
\ref{3},  we have seen that the energy curve shows a minimum for
each relevant density. This behavior indicates that for each
density there is a metastable state. We have also seen that  as
the  density increases, this metastable state is shifted to lower
values of the polarization parameter. Therefore, we can conclude
that the metastable state disappears at high densities. We have
also found that at high densities, the system becomes nearly
identical with the unpolarized case. These results agree with
those of reference \cite{6}.
In Fig. \ref{4},  we have plotted the total energy density of the
spin polarized SQM  versus the number density in the presence of
magnetic field. We have seen that the total energy increases by
increasing the density. We have found that the  energy density of
the spin polarized SQM in the presence of magnetic field is nearly
identical with that of the unpolarized case which has been
clarified in panel (b) of  Fig. \ref{4}. As we will see in the
next paragraph, this is due to the fact that the polarization
parameter in the presence of magnetic field is very small,
especially at high densities.

\begin{figure}
\includegraphics[width=14cm]{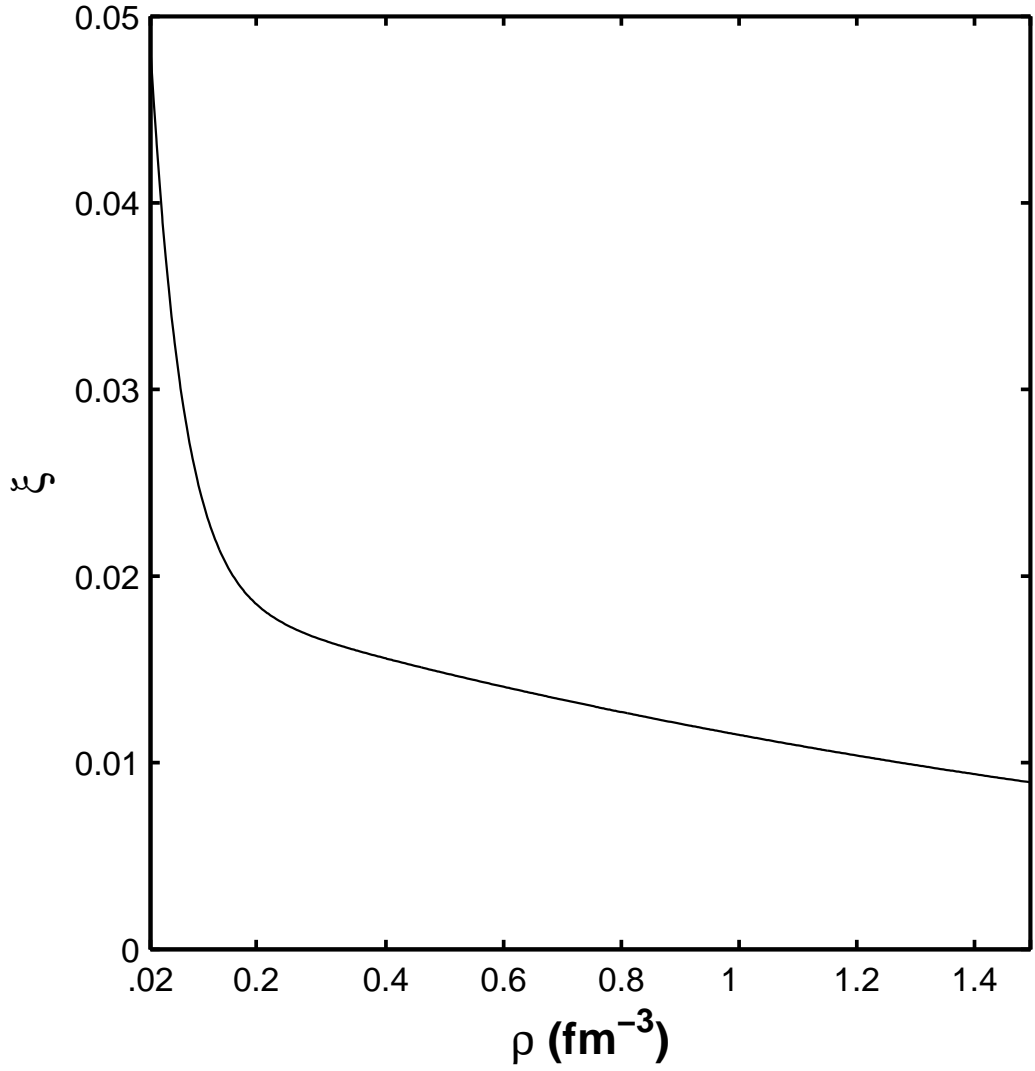}
\caption{The polarization parameter ($\xi$) corresponding to the
minimum points of the energy density versus the density ($\rho$)
at $B=5\times10^{18}\ G$.} \label{5}
\end{figure}

\begin{figure}
\includegraphics[width=14cm]{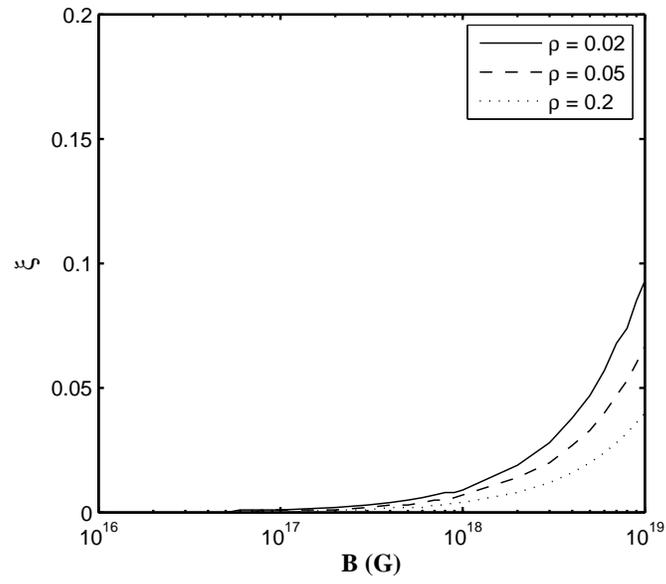}
\caption{The polarization parameter ($\xi$) corresponding to the
minimum points of the energy density versus the magnetic field
($B$) for different values of density ($\rho$).}\label{6}
\end{figure}

In Fig. \ref{5}, we have presented the polarization parameter
corresponding to the minimum point of energy density as a function
of the number density at $B=5\times10^{18}\ G$. We see that the
polarization parameter decreases by increasing the number density.
From Fig. \ref{5}, it can be seen that for $\rho<0.2\ fm^{-3}$,
the decreasing rate of polarization versus density is
substantially higher than for $\rho>0.2\ fm^{-3}$.
In Fig. \ref{6}, we have shown the polarization parameter versus
the magnetic field for different values of the number density. For
each density, we can see that  the polarization increases by
increasing the magnetic field. This figure also shows that the
increasing rate of polarization versus magnetic field increases by
increasing density.

\begin{figure}
\includegraphics[width=14cm]{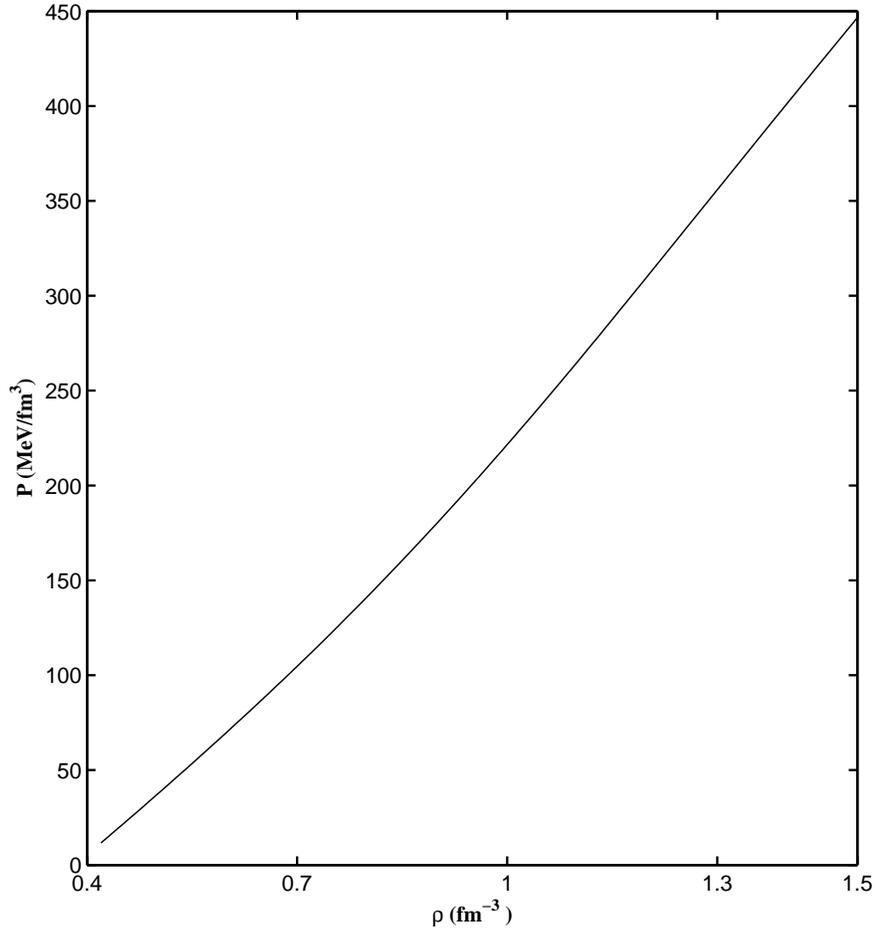}
\caption{ The pressure ($P$) versus density ($\rho$) for spin
polarized SQM at $B=5\times10^{18}\ G$. }\label{7}
\end{figure}

We have also calculated EoS of spin polarized SQM in the presence
of the magnetic field, where the contribution of magnetic pressure
($\frac{B^{2}}{8\pi}$) should be added to Eq. (\ref{eos}) in which
the total energy density is obtained from Eq. (\ref{equa121}). In
Fig. \ref{7}, we have plotted EoS of spin polarized SQM where the
magnetic field is switched on. We have found that this EoS is
nearly identical with that of the unpolarized case. This is due to
the fact that polarization at minimum of energy is very low,
especially at high densities.

\begin{figure}
\includegraphics[width=14cm]{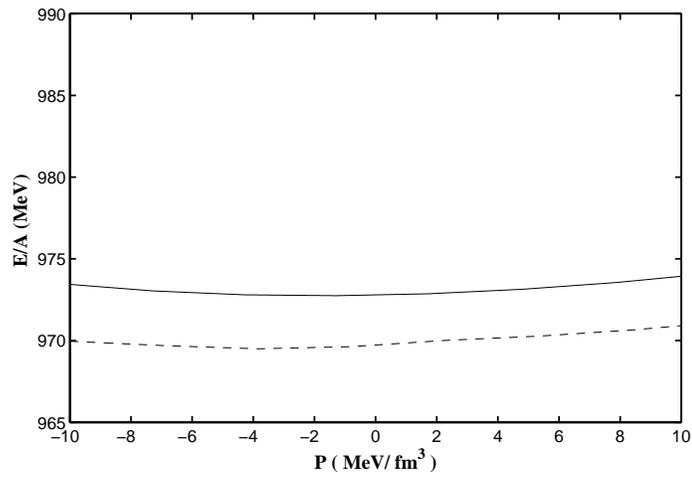}
\caption{The energy per baryon versus the pressure ($P$) for spin
polarized SQM at $B=0$ (full curve) and $B= 5\times 10^{18}\ G$
(dashed curve). }\label{8}
\end{figure}

In Fig. \ref{8}, we have plotted the energy per baryon ($E/A$) for
the spin polarized SQM as a function of  pressure  at $B=5 \times
10^{18}\ G$. Our results for the case of SQM in the absence of
magnetic field ($B=0$) are also given for comparison. We have seen
that the zero point of pressure in the presence of magnetic field
has a lower $E/A$ compared to the case of SQM in the absence of
magnetic field ($B=0$). This indicates that, in the presence of
magnetic field, the spin polarized SQM is more stable than that in
the absence of magnetic field.


\section{Structure of the spin polarized SQS}
The gravitational mass ($M$) and radius ($R$) of  compact stars
are of special interests in astrophysics. In this section, we
calculate the structure properties of spin polarized SQS and
compare the results of this calculation with those of the
unpolarized case. Using the EoS of spin polarized SQM, We can
obtain $M$ and $R$  by numerically integrating the general
relativistic equations of hydrostatic equilibrium,
Tolman-Oppenheimer-Volkoff (TOV) equations, which are as follows
\cite{9},

\begin{eqnarray}
\frac{dm}{dr} &=& 4\pi r^{2}\varepsilon(r),\nonumber \\
 \frac{dP}{dr}&=&
-\frac{Gm(r)\varepsilon(r)}{r^{2}}\left(1+\frac{P(r)}{\varepsilon(r)c^{2}}
\right)\left(1+\frac{4\pi
r^{3}P(r)}{m(r)c^{2}}\right)\left(1-\frac{2Gm(r)}{c^{2}r}
\right)^{-1},\label{tov1}
\end{eqnarray}
where $\varepsilon(r)$ is the energy density, $G$ is the
gravitational constant, and
\begin{equation}
m(r) =\int_0^r 4\pi r'^2\varepsilon (r')dr'\label{tov2}
\end{equation}
has the interpretation of the mass inside radius $r$. By selecting
a central energy density $\varepsilon_{c}$, under the boundary
conditions $P(0)=P_{c}$,  $m(0)=0$, we integrate the TOV equation
outwards to a radius $r=R$, at which $P$ vanishes. This yields the
radius $R$ and mass $M=m(R)$ \cite{9}.

Our results for the structure of spin polarized SQS in the absence
and presence of the magnetic field are given separately in two
following sections.

\subsection{Structure of the spin polarized SQS in the absence of magnetic field}

\begin{figure}
\includegraphics[width=14cm]{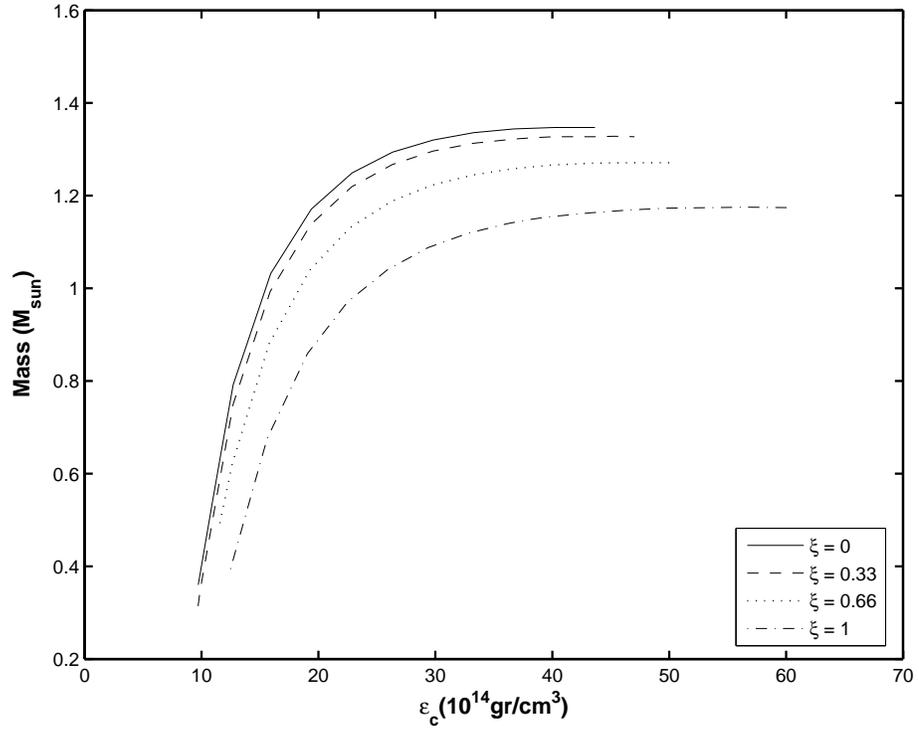}
\caption{The gravitational mass of spin polarized SQS versus the
central density ($\varepsilon_c$) for different values of the
polarization parameter ($\xi$) in the absence of magnetic field.
}\label{9}
\end{figure}

\begin{figure}
\includegraphics[width=14cm]{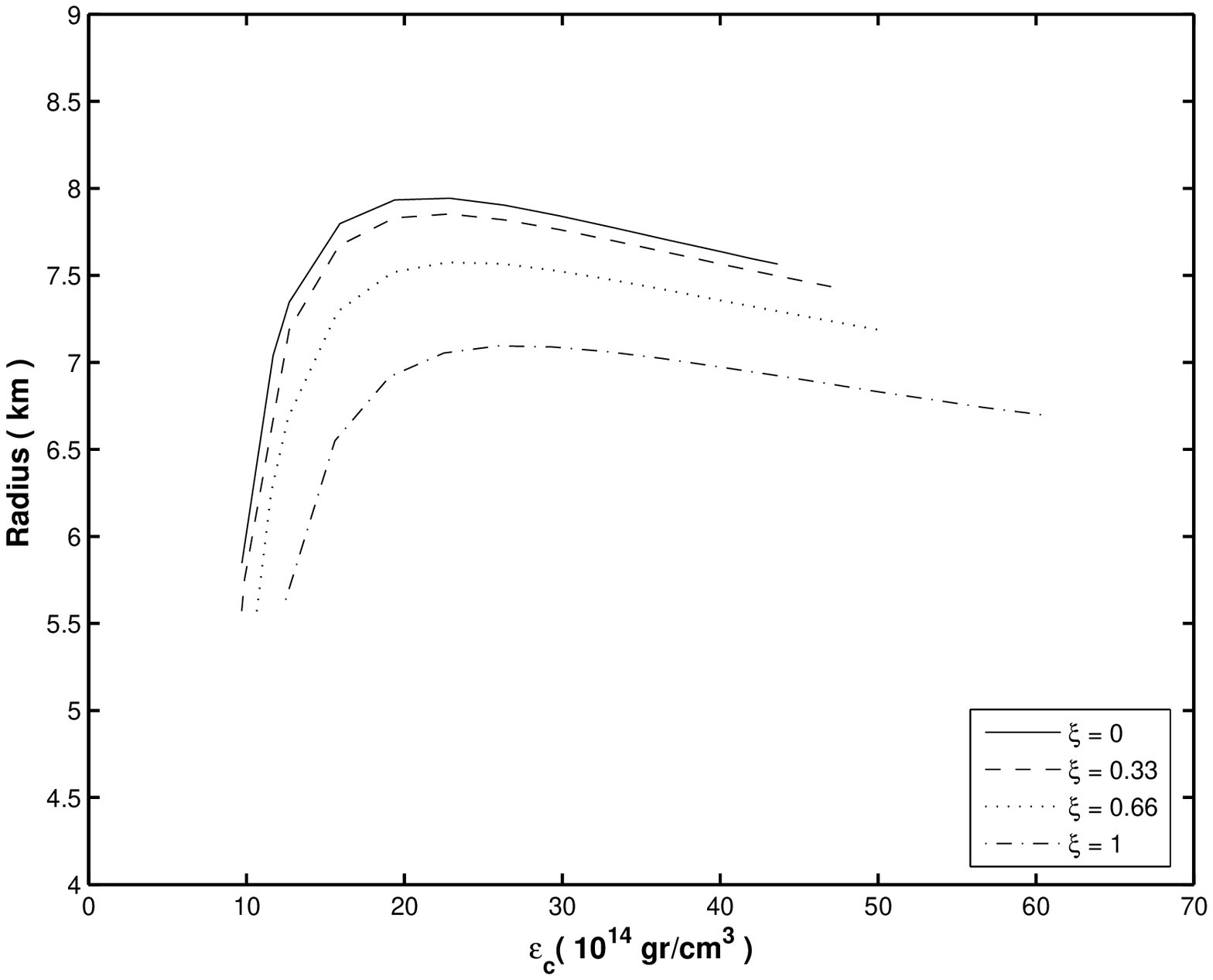}
\caption{As Fig. \ref{9} but for the radius of spin polarized SQS.
}\label{10}
\end{figure}

\begin{figure}
\includegraphics[width=14cm]{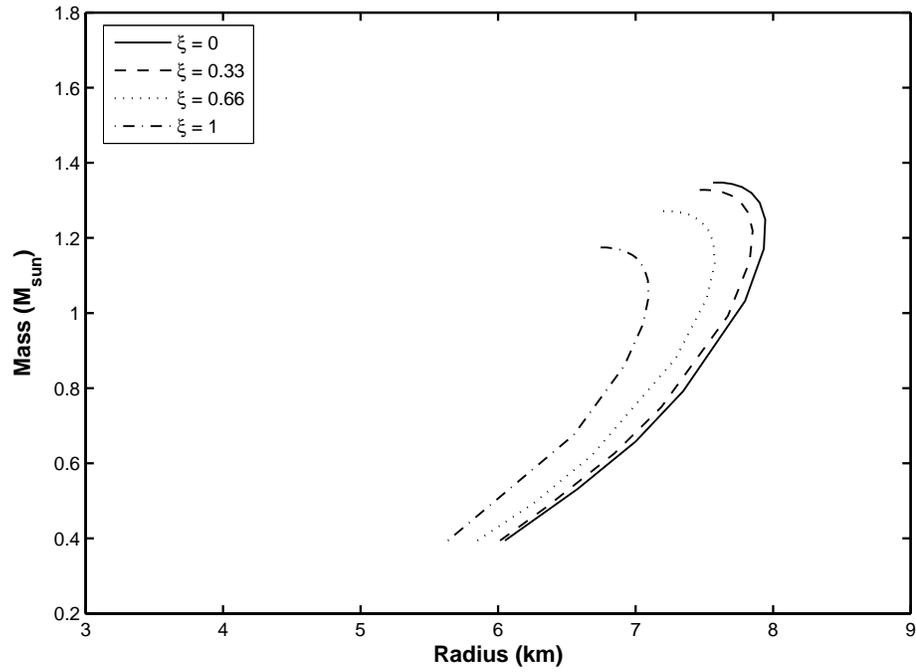}
\caption{The mass-radius relation for spin polarized SQS in the
absence of magnetic field at different values of the polarization
parameter ($\xi$). }\label{11}
\end{figure}

In Figs. \ref{9} and \ref{10}, we have plotted the gravitational
mass and radius of the spin polarized SQS in the absence of
magnetic field versus the central energy density
$(\varepsilon_{c})$ for different values of the polarization
parameter ($\xi$).
From these figures, we see that for each central density, the mass
and radius of SQS decrease by increasing the polarization
parameter. This is due to the fact that by increasing the
polarization parameter, the pressure of spin polarized SQM
increases, which leads to the stiffer equation of state for this
system (Fig. \ref{2}). Figs. \ref{9} and \ref{10} show that for a
given polarization parameter, the gravitational mass and radius of
SQS increase by increasing the central density. From Fig. \ref{9},
it can be seen that the gravitational mass of SQS reaches a
limiting value called the maximum mass.
In Fig. \ref{11}, we have plotted our results for the
gravitational mass of spin polarized SQS as a function of the
radius (mass-radius relation) in the absence of magnetic field.
For this system, we see that the gravitational mass increases by
increasing the radius.
It is seen that the rate of increasing mass versus radius
increases by increasing the polarization.
In Table \ref{1}, the maximum mass ($M_{max}$) and the
corresponding radius ($R$) of spin polarized SQS have been given
for different values of the polarization parameter ($\xi$) in the
absence of magnetic field. We can see that both maximum mass and
the corresponding radius decrease by increasing $\xi$.
This shows that increasing polarization leads to a more stable
SQS.

\subsection{Structure of the spin polarized SQS in the presence of magnetic field  }
\begin{figure}
\includegraphics[width=14cm]{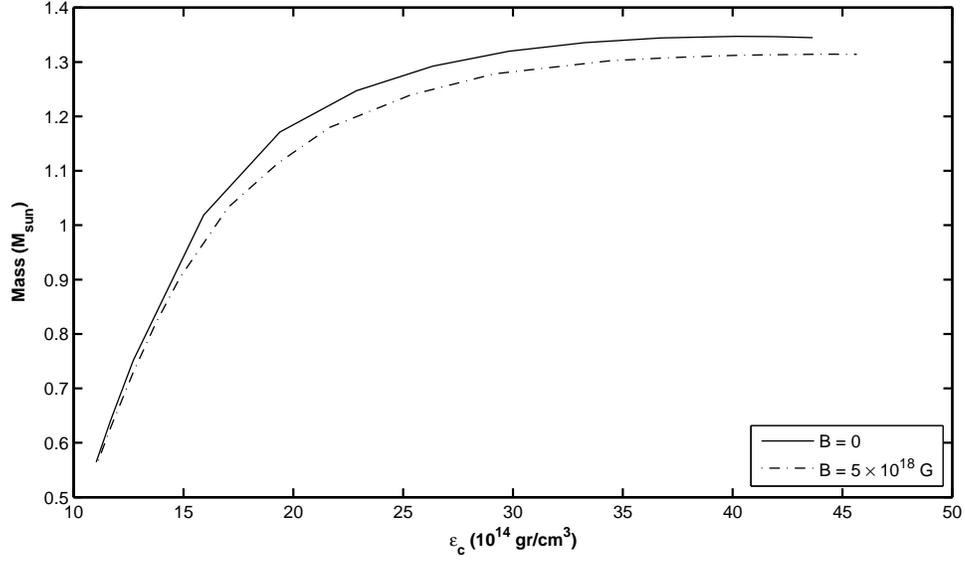}
\caption{The gravitational mass versus the central density
($\varepsilon_c$) for the spin polarized SQS at $B=0$ and $B=
5\times 10^{18}\ G$. }\label{meb}
\end{figure}

\begin{figure}
\includegraphics[width=14cm]{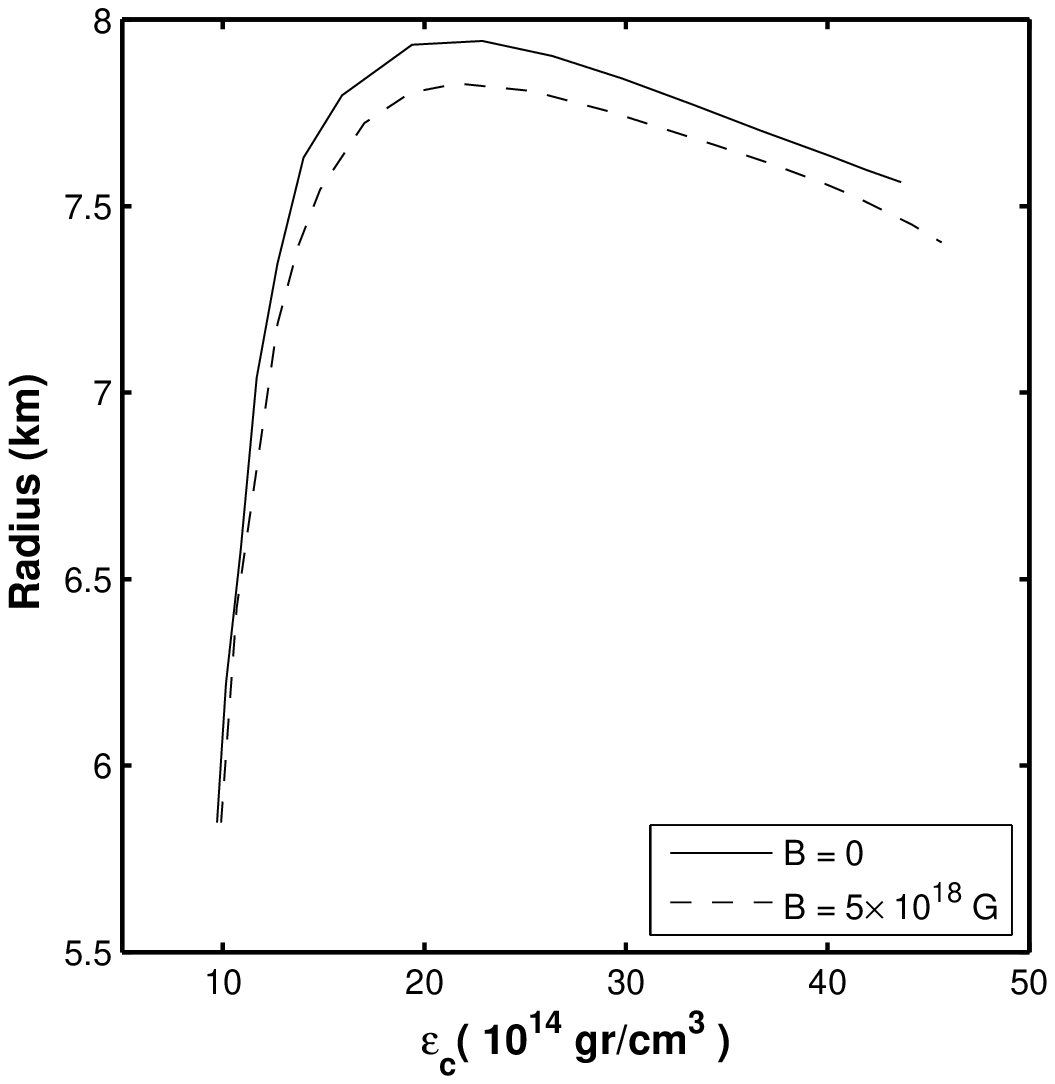}
\caption{As Fig. \ref{meb} but for the radius of spin polarized
SQS. }\label{reb}
\end{figure}

In this section, we present our calculations for the structure of
SQS in the presence of the magnetic field.
It should be noted that the strong magnetic field changes the
spherical symmetry of the system. However, for the magnetic fields
less than $10^{19}\ G$, this effect is negligible \cite{gonzalez,
perez}, therefore, we can solve the TOV equations using a
spherical metric, which leads to Eq. (\ref{tov1}).
Our results for the gravitational mass and radius of the spin
polarized SQS in the presence of magnetic field versus the central
energy density $(\varepsilon_{c})$ have been shown in Figs.
\ref{meb} and \ref{reb}, respectively. In these figures, our
results for the unpolarized case of SQS ($B=0$) are also given for
comparison.
Figs. \ref{meb} and \ref{reb} show that for all values of central
density, the mass and radius of SQS decrease when the magnetic
field is switched on.
From Fig. \ref{meb}, we see that as the central density increases,
the gravitational mass of SQS increases and finally reaches a
limiting value (maximum mass).
In Table \ref{2}, we have given the maximum mass  and the
corresponding radius  of SQS for two cases $B=0$ (unpolarized SQS)
and $B=5\times10^{18}\ G$.
It is shown that the presence of magnetic field leads to lower
values for both maximum mass and the corresponding radius of SQS
showing a more stable SQS compared to the unpolarized SQS.
%
\section{Summary and Conclusions}
We have studied the spin polarized strange quark matter (SQM) for
both cases in the absence and presence of magnetic field. We have
calculated some of the bulk properties of this system such as the
energy, equation of state (EoS), and polarization.
We have shown that the energy of spin polarized SQM in the absence
of magnetic field increases by increasing polarization.
Calculation of energy in the presence of magnetic field shows
that  for each density, there is a minimum point for the energy of
SQM showing a metastable state.
We have seen that the EoS of spin polarized SQM becomes stiffer as
the polarization increases.
We have also seen that the spin polarized SQM in the presence of
magnetic field is more stable than the unpolarized SQM.
The structure properties of spin polarized strange quark star
(SQS) have been also calculated in the absence and presence of the
magnetic field.
We have seen that for each central density, the mass and radius of
spin polarized SQS decrease by increasing polarization. We have
also seen that both maximum mass and the corresponding radius of
this system decrease by increasing polarization.
We have indicated that in the presence of magnetic field, the
maximum mass and the corresponding radius of the polarized SQS get
lower values than those of unpolarized SQS. Therefore, we can
conclude that the presence of magnetic field leads to a more
stable SQS compared to the unpolarized SQS.

Our results for the maximum mass and radius of SQS (Tables \ref{1}
and \ref{2}) are consistent with those observed for the object SAX
J1808.4-3658 \cite{li}. We can conclude that this object is a good
candidate for SQS.

One of the other astrophysical implications of our results is
calculation of the surface redshift $(z_{s})$ of SQS. This
parameter is of special interest in astrophysics and can be
obtained from the mass and radius of the star using the following
relation \cite{10},
\begin{eqnarray}
    z_{s} =(1-\frac{2GM}{Rc^{2}})^{-\frac{1}{2}}-1.
\end{eqnarray}
Our results corresponding to the maximum mass and radius of SQS
lead to $z_{s}=0.45\ m\,s^{-1} $ in the absence of magnetic field
and $z_{s}=0.44 \ m\,s^{-1} $ for the magnetic field $B=5\times
10^{18} \ G$. This indicates that the presence of magnetic field
leads to the (nearly) lower values for the surface redshift.

\section*{Acknowledgements} {This work has been supported by Research
Institute for Astronomy and Astrophysics of Maragha. We wish to
thank Shiraz University and Tafresh University Research Councils.
One of us (A. R. Peivand) also wishes to thank M. Mirza.}

\newpage
\begin{table}[ht]
\begin{center}
\footnotesize\rm \caption{\footnotesize\rm Maximum gravitational
mass ($M_{max}$) and the corresponding radius ($R$) of the spin
polarized SQS for different values of the polarization parameter.
} \footnotesize\rm
\begin{tabular}{@{}l@{}c@{}c@{}c@{}}

\\\hline\hline
\,\,\,\,\,\,\,$\mathbf{Star}$ & \,\,\,$\mathbf{M_{max}\
(M_{\odot})}$\,\, & \,\, $\mathbf{R\ (km)}$\\\hline\hline
\footnotesize  Unpolarized SQS $(\xi=0)$& 1.35&7.6 \\
\hline \footnotesize Polarized SQS  (${\xi=0.33}$) &1.33&7.5
\\\hline
\footnotesize Polarized SQS (${\xi=0.66}$) &1.27&7.2 \\
\hline
\footnotesize Polarized SQS (${\xi=1}$) &1.17&6.7 \\

\hline\hline\\
\end{tabular}
\end{center}
\end{table}
\newpage

\begin{table}[ht]
\begin{center}
\footnotesize\rm
 \caption{\footnotesize\rm Maximum gravitational
mass ($M_{max}$) and the corresponding radius ($R$) of SQS for
$B=0$ and $5\times10^{18}\ G$. } \footnotesize\rm
\begin{tabular}{@{}l@{}c@{}c@{}c@{}}

\\\hline\hline
\,\,\,\,\,\,\,$\mathbf{Star}$ & \,\,\,$\mathbf{M_{max}\
(M_{\odot})}$\,\, & \,\, $\mathbf{R\ (km)}$\\
\hline\hline
\footnotesize  Unpolarized SQS $(B=0)$& 1.35&7.6\\
\hline
\footnotesize Polarized SQS (${B=5\times10^{18}G\, }$) & 1.31&7.4\\  \hline\hline\\
\end{tabular}
\end{center}
\end{table}


\newpage
\newpage
\newpage
\newpage
\newpage
\newpage
\newpage
\newpage
\newpage

\newpage
\newpage
\newpage
\newpage

\newpage
\begin{thebibliography}{99}

\bibitem{haensel} Haensel P., Potekhin A. Y., Yakovlev D. G.,
2007, \emph{Neutron Stars 1}, New York: Springer.

\bibitem{glendening} Glendenning N. K., 2000, \emph{Compact Stars:
Nuclear Physics, Particle Physics, and General Relativity}, New
York: Springer.

\bibitem{weber} Weber F., 1999, \emph{Pulsars as Astrophysical Laboratories
for Nuclear and Particle Physics}, Bristol: IOP Publishing.

\bibitem{10}  Camenzind M., 2007, \emph{Compact Objects
in Astrophysics: White Dwarfs, Neutron Stars and Black Holes} ,
Springer.

\bibitem{a} Itoh N., 1970, {Prog. Theor. Phys.} {\bf 44}, 291.
\bibitem{b}  Bodmer A. R., 1971, {Phys. Rev. } {\bf D 4}, 1601.
\bibitem{collins} Collins J. C., Perry M. G., 1975, {Phys. Rev.
Lett.} {\bf 34}, 1353.
\bibitem{c}  Witten E., 1984, {Phys. Rev. } {\bf D 30}, 272.

\bibitem{d}  Alcock C., Farhi E.,  Olinto A.,  1986, {Astrophy. J.}
{\bf 310}, 261.
\bibitem{e}  Haensel P.,  Zdunik J. L., Schaeffer R., 1986,
{Astron. Astrophys.} {\bf 160}, 121.
\bibitem{f}  Kettner C., Weber F.,  Weigel M. K.,  Glendenning N.
K., 1995, {Phys. Rev.} {\bf D 51}, 1440.

\bibitem{2} Bhattacharyya A. et al., 2006, Phys. Rev. {\bf C 74},
065804.


\bibitem{prakash} Prakash M., Lattimer J. M., Steiner A. W., Page
D., 2003, {Nucl. Phys.} {\bf A 715}, 835.

\bibitem{kouv1} Kouveliotou C. et al., 1999, Astrophys. J. {\bf 510},
L115 .
\bibitem{kouv2} Kouveliotou C. et al., 1998, Nature {\bf 393}, 235.

\bibitem{chak}  Ghosha T., Chakrabarty S., 2001, Phys. Rev.  {\bf D 63}, 043006.

\bibitem{nurafshan}  Bordbar G. H.,  Nourafshan M. and
Khosropour B., 2009, Iranian J. Phys. Res. {\bf 9}, 237 .

\bibitem{zamani} Bordbar G. H.,  Poostforush A. and  Zamani A.,
Astrophys. (2011) accepted for publication.

\bibitem{chodos}  Chodos A. et al., 1974, {Phys. Rev.} {\bf D 9}, 3471 .


\bibitem{6} Pal K., Biswas S.,  Dutt-Mazumder A. K., 2009, Phys. Rev. {\bf C 79},
015205.

\bibitem{9} Shapiro Stuart L. and Teukolsky Saul. A., 1983,
\emph{Black Holes, White Dwarfs, and Neutron
Stars: The Physics of Compact Objects}, NewYork:
Wiley-Interscience, First edition.



\bibitem{gonzalez}  Gonzalez Felipe R., Perez Martinez A., 2009, J. Phys. {\bf G 36},
075202.
\bibitem{perez}  Perez Martinez A., Gonzalez Felipe R., Manreza Paret
D., 2010, arXiv:1001.4038.


\bibitem{li}  Li X.-D et al, 1999, Phys.Rev.Lett. {\bf 83},
3776-3779.


\end{thebibliography}
\end{document}